\documentclass[epj,dvips]{webofc}
\usepackage[varg]{txfonts}   % Web of Conferences font
\providecommand{\tabularnewline}{\\}
\newcommand{\aap}{A\&A}

\newcommand{\apj}{ApJ}
\newcommand{\apjl}{\apj}

\newcommand{\apjs}{ApJS}

\newcommand{\dd}[2][]{\frac{d #1}{d #2}}                  % derivative
\woctitle{Physics at the Magnetospheric Boundary}
\begin{document}
\title{Boundary Between Stable and Unstable Regimes of Accretion}

\author{A.~A. Blinova\inst{1}\fnsep\thanks{\email{alisablinova@gmail.com}}, R.~V.~E. Lovelace\inst{1},
M.~M. Romanova\inst{1}}

\institute{Department of Astronomy, Cornell University, Ithaca, NY
14853-6801,
  USA}

\abstract{%
We investigated the boundary between stable and unstable regimes
of accretion and its dependence on different parameters.
Simulations were performed using a ``cubed sphere" code with high
grid resolution (244 grid points in the azimuthal direction),
which is twice as high as that used in our earlier studies. We
chose a very low viscosity value, with alpha-parameter
$\alpha$=0.02. We observed from the simulations that the boundary
strongly depends on the ratio between magnetospheric radius $r_m$
(where the magnetic stress in the magnetosphere matches the matter
stress in the disk) and corotation radius $r_{\rm cor}$ (where the
Keplerian velocity in the disk is equal to the angular velocity of
the star). For a small misalignment angle of the dipole field,
$\Theta=5^\circ$, accretion is unstable if $r_{\rm cor}/r_m>1.35$,
and is stable otherwise. In cases of a larger misalignment angle
of the dipole, $\Theta=20^\circ$, instability occurs at slightly
larger values, $r_{\rm cor}/r_m>1.41$.}
\maketitle
\vspace{-0.2cm}\section{Introduction} \label{intro}

\begin{figure}[!ht]
\centering
\includegraphics[width=8cm,clip]{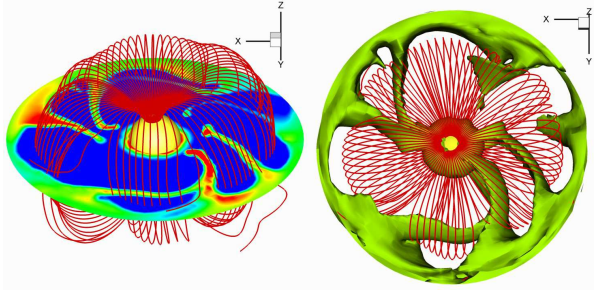}
\vspace{-0.3cm}\caption{3D views of accretion through
Rayleigh-Taylor instability. \textit{Left panel}: sample field
lines and equatorial density distribution. \textit{Right panel}: a
3D top-down view.}
\label{3d-unstable}       % Give a unique label
\end{figure}

Magnetospheric accretion occurs in different types of stars,
including Classical T Tauri stars (CTTSs) \cite{bouvier2007},
magnetized cataclysmic variables \cite{warner2004}, and accreting
millisecond pulsars \cite{vanderklis2006}. Matter can accrete to
magnetized stars in the stable regime, where ordered funnel
streams are formed and subsequently flow to the magnetic poles, or
in the unstable regime (see Fig.~\ref{3d-unstable}), where matter
penetrates through the magnetosphere in several unstable tongues
due to the magnetic Rayleigh-Taylor instability
\cite{kulkarni2008,arons1976,romanova2008}.

Stars accreting in stable and unstable regimes have different
observational properties. In the stable regime, matter of the
funnel streams falls to the star and forms two antipodal hot spots
on its surface, and the expected light-curve is nearly sinusoidal.
In the unstable regime, several hot spots form per rotational
period of the inner disk, and the light-curve may look stochastic,
with several peaks per period of rotation
\cite{kulkarni2008,romanova2008}. Such stochastic light-curves are
also expected to be observed in the spectral lines, particularly
in their red-shifted components \cite{kurosawa2013}. Therefore, it
is important to know which stars are expected to be in the stable
or unstable regimes of accretion, and to derive the boundaries
between these two regimes.

The main research objective was to investigate the boundary
between stable and unstable regimes of accretion in simulations
with a high grid resolution and at a low viscosity parameter,
$\alpha=0.02$, in the disk.

%and its dependence on the stellar period (related to corotation
%radius $r_{cor}$) and strength of the dipole magnetic moment $\mu$
%(related to magnetospheric radius $r_m$). In addition, the effects
%of varying the tilt of the magnetic dipole on the boundary between
%stable and unstable regimes of accretion were explored.

%Figure 1%

\vspace{-0.3cm}\section{Method} \label{method}

To investigate the boundary between stable and unstable regimes of
accretion, we performed multiple 3D MHD (magnetohydrodynamic)
simulation runs using a Godunov-type numerical code, written in a
``cubed sphere" coordinate system rotating with the star
\cite{koldoba2002}. Simulations were performed using a high grid
resolution (244 grid points in the azimuthal direction), which is
twice as high as that used in our earlier studies. We considered
accretion from an $\alpha$-disk \cite{shakura1973} with a small
$\alpha$-parameter, $\alpha=0.02$.

\vspace{-0.3cm}
\section{Results} \label{results}

In our simulations we varied the dipole magnetic moment ($\mu$)
from $\mu=0.05-3$, and the corotation radius from $r_{\rm
cor}=1.2-3$ (in dimensionless units). We found that the boundary
strongly depends on the ratio of corotation radius $r_{\rm cor}$
to magnetospheric radius $r_m$. $r_m$ was obtained from the
simulations. We ran simulations for two different misalignment
angles of the dipole, $\Theta=5^\circ$ and $\Theta=20^\circ$.

%%%%%%% Figure %%%%%%%%%%%%%%%%%%%%%%%%%%%%
%\begin{figure*}
%  \begin{center}
%    \includegraphics[clip,width=0.84\textwidth]{fig02.eps}
%  \end{center}
%\vspace{-0.5cm} \caption{The } \label{model-profiles}
%\end{figure*}
%%%%%%%%%%%%%%%%%%%%%%%%%%%%%%%%%%%%%%%%%%%%%

\begin{figure}[!ht]
\centering
\includegraphics[width=12cm,clip]{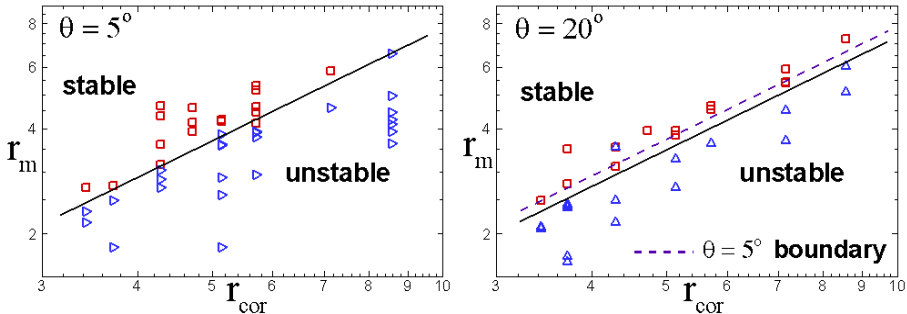}
\vspace{-0.3cm}\caption{The boundary between stable and unstable
regimes of accretion in the parameter space of $r_m$ and $r_{\rm
cor}$ for two different misalignment angles of the dipole. The
units are given in stellar radii. \textit{Left panel}: the
boundary line for $\Theta=5^\circ$. \textit{Right panel}:
superposition of the $\Theta=5^\circ$ boundary line with the
$\Theta=20^\circ$ data set and line.}
\label{Rm-Rc-Theta5-20-final-r-star}       % Give a unique label
\end{figure}

\begin{figure}
\centering
\includegraphics[width=8cm,clip]{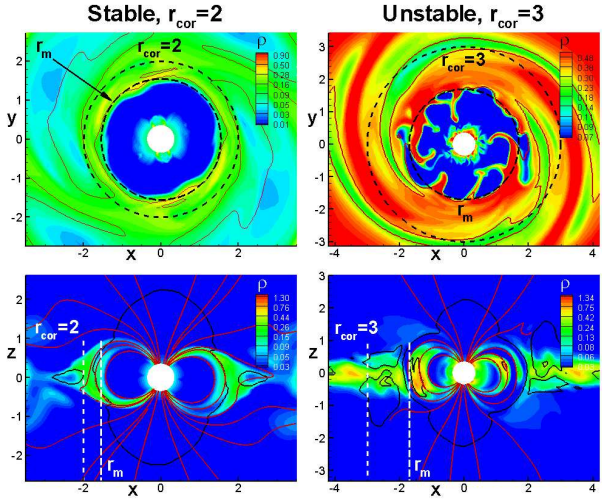}
\vspace{-0.3cm}\caption{Examples of slices of density distribution
in stable and unstable cases for $\Theta=5^\circ$, where dipole
moment $\mu=2$ and only $r_{\rm cor}$ is varied. \textit{Left
panels}: xy-slice (top) and xz-slice (bottom) show the case of
$r_{\rm cor}=2$, where accretion is stable. \textit{Right panels}:
accretion becomes unstable when $r_{\rm cor}=3$. Here, the units
are dimensionless, with $r_{\rm star}=0.35$.}
\label{d2-stable-unstable-final}       % Give a unique label
\end{figure}

For a small misalignment angle of the dipole, $\Theta=5^\circ$,
accretion is unstable if  $r_{\rm cor}/r_m>1.35$, and is stable
otherwise (see Fig.~\ref{Rm-Rc-Theta5-20-final-r-star}, left
panel). That is, instability occurs more easily for larger
corotation radii and smaller magnetospheric radii. Alternatively,
when $r_m$ approaches $r_{\rm cor}$, accretion becomes stable. For
example, when we take two cases with the same magnetic moment,
$\mu=2$, but two different corotation radii, $r_{\rm cor}=2$ and
$r_{\rm cor}=3$, we find that for the same moment in time the
former case is stable, while the latter case is unstable (see
Fig.~\ref{d2-stable-unstable-final}). For a larger misalignment
angle, $\Theta=20^\circ$, instability occurs at slightly larger
values, $r_{\rm cor}/r_m>1.41$ (see
Fig.~\ref{Rm-Rc-Theta5-20-final-r-star}, right panel). This is
probably due to the fact that at larger misalignment angles the
magnetic poles are closer to the accretion disk plane, which makes
funnel stream formation more favorable than the Rayleigh-Taylor
instability.

%Figure 2%

%Figure 3%

It should also be noted that not all cases were completely stable
or completely unstable during a single simulation run. We broke
these "intermediate" cases up into several stable and unstable
parts at different moments in time, and recorded their
corresponding magnetospheric radii.

\begin{figure}
\centering
\includegraphics[width=7cm,clip]{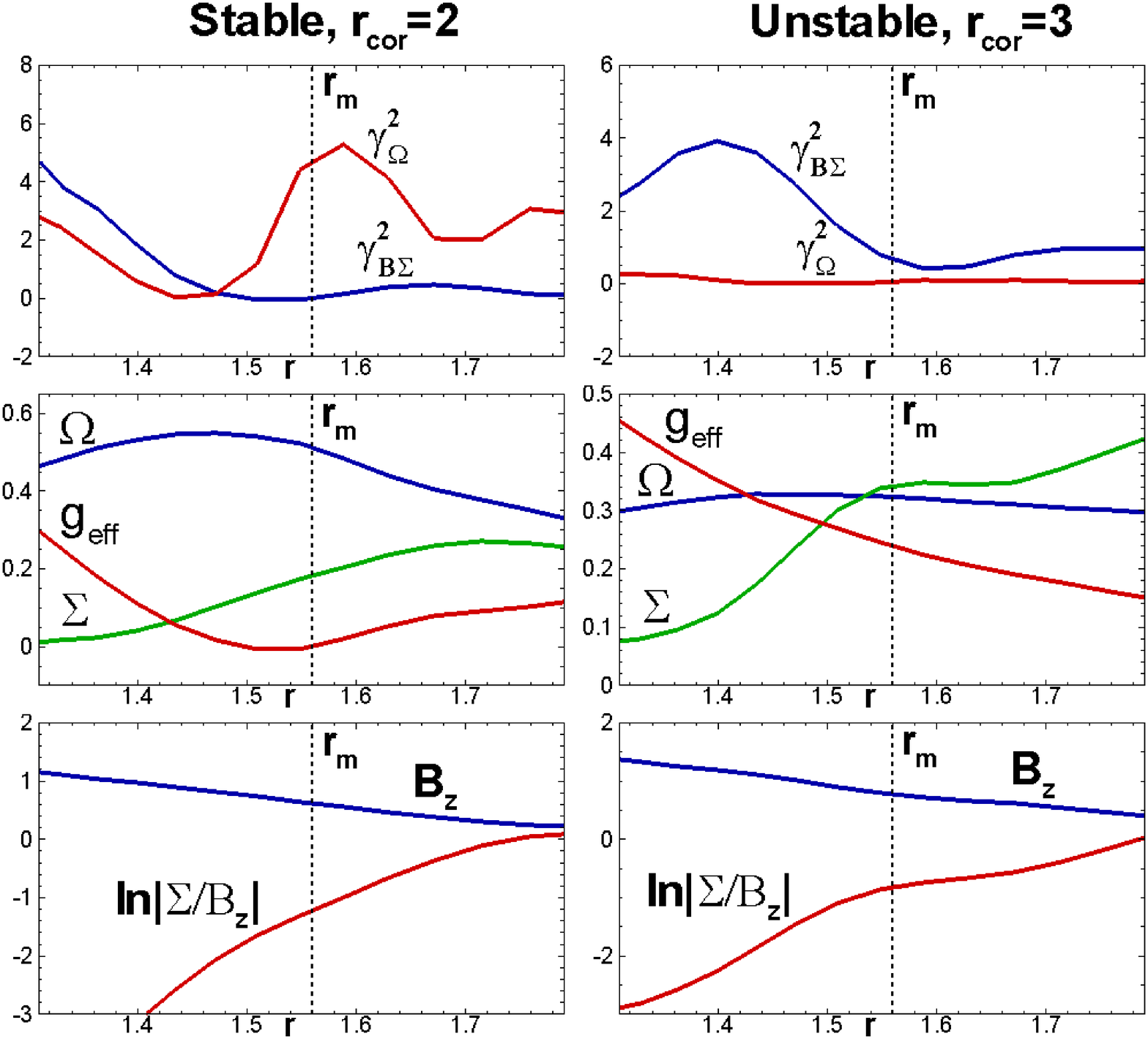}
\vspace{-0.3cm}\caption{Radial distribution of different
parameters in the disk and terms of the Spruit criterion in the
vicinity of $r_m$. Dipole magnetic moment $\mu=2$ (in
dimensionless units). $\Theta=5^\circ$. Accretion is stable for
the smaller $r_{\rm cor}$, $\gamma_{B\Sigma}^2 < \gamma_\Omega^2$
(\textit{left panel}), and unstable for the larger $r_{\rm cor}$,
$\gamma_{B\Sigma}^2 > \gamma_\Omega^2$ (\textit{right panel}).}
\label{sigma-av-d2-v3}       % Give a unique label
\end{figure}

We compared our simulation results with the analytical criterion
for the existence of instability by Spruit et al.
~\cite{spruit1995}:

\begin{equation}
\gamma_{B\Sigma}^2 \equiv g_{\rm eff} \left| \dd{r} \ln
\frac{\Sigma}{B_z} \right| > 2 \left( r \dd[\Omega]{r} \right)^2
\equiv \gamma_\Omega^2, \label{eq:spruit}
\end{equation}
where $\Sigma$ is the surface density in the disk; $B_z$ is the
z-component of the magnetic field; $\Omega$ is the angular
velocity of the disk; $g_{\rm eff} \equiv g - \Omega^2r$ is the
effective gravitational acceleration ($g_{\rm eff}$ is positive if
the acceleration is radially inwards). In other words, the
${\Sigma}/{B_z}$ and $g_{\rm eff}$ terms should increase fast
enough with $r$ to overcome the effects of the shear term of
angular velocity, $\gamma_\Omega^2$, which has a stabilizing
effect on the accretion disk. The majority of our cases for both
misalignment angles are consistent with the Spruit criterion. That
is, at the magnetospheric radius (where the unstable perturbations
occur), $\gamma_{B\Sigma}^2 > \gamma_\Omega^2$ for the unstable
cases, while $\gamma_{B\Sigma}^2 < \gamma_\Omega^2$ for the stable
cases, which is consistent with the analytical prediction. See
Fig.~\ref{sigma-av-d2-v3} for an example of the Spruit criterion
applied to two cases with the same dipole magnetic moment,
$\mu=2$, but different corotation radii.

%Figure 4%

Note that these simulations are different from those of
\cite{kulkarni2008,romanova2008} in that we used $\alpha=0.02$ in
all simulation runs (which approximately corresponds to constant
accretion rate), while in \cite{kulkarni2008,romanova2008} the
$\alpha$-parameter (and hence the accretion rate) varied. The
comparison of our results with previous works shows that
instability occurs more readily when using a finer grid.

\vspace{-0.3cm}\section{Summary} \label{summary}

The boundary between stable and unstable regimes of accretion
strongly depends on the ratio $r_{\rm cor}/r_m$. Accretion is
slightly more stable for larger misalignment angles $\Theta$, with
$r_{\rm cor}/r_m$ having a slightly higher value. The results of
our simulations are also consistent with the Spruit criterion for
instability.

\vspace{-0.2cm}\section*{Acknowledgments} We thank the conference
organizers, particularly Dr. E. Bozzo for organizing an excellent
conference. Resources supporting this work were provided by the
NASA High-End Computing (HEC) Program through the NASA Advanced
Supercomputing (NAS) Division at Ames Research Center and the NASA
Center for Computational Sciences (NCCS) at Goddard Space Flight
Center.  The research was supported by NASA grant NNX11AF33G and
NSF grant AST-1008636.

%
% BibTeX or Biber users please use (the style is already called in the class, ensure that the "woc.bst" style is in your local directory)
% \bibliography{name or your bibliography database}
%
%\bibliography{local}

\vspace{-0.2cm}\begin{thebibliography}{}
\bibitem{bouvier2007}
Bouvier J.,  Alencar S. H. P., Harries T. J., Johns-Krull C. M.,
Romanova M. M., \textit{Protostars and Planets V}, Eds. Reipurth
B., Jewitt D., Keil K. (University of Arizona Press, Tucson, 2007)
479
\bibitem{warner2004}
Warner B., PASP \textbf{116}, 115 (2004)
\bibitem{vanderklis2006}
van der Klis M., \textit{Compact Stellar X-Ray Sources}, Eds.
Lewin W. H. G. and van der Klis M. (Cambridge Univ. Press,
Cambridge, 2006) 39
\bibitem{kulkarni2008}
Kulkarni A. K. and Romanova M. M., MNRAS \textbf{386}, 673 (2008)
\bibitem{arons1976}
Arons J. and Lea S. M., ApJ \textbf{207}, 914 (1976)
\bibitem{romanova2008}
Romanova M. M., Kulkarni A. K., Lovelace R. V. E., ApJ
\textbf{673}, L171 (2008)
\bibitem{kurosawa2013}
Kurosawa R. and Romanova M. M., MNRAS in press, e-print
arXiv:1307.3639 (2013)
\bibitem{koldoba2002}
Koldoba A. V., Romanova M. M., Ustyugova G. V., Lovelace R. V. E.,
ApJ \textbf{576}, L53 (2002)
%Book Author, \textit{Book title} (Publisher, place, year) page
%numbers
% etc
\bibitem{shakura1973}
Shakura N. I. and Sunyaev R. A., A\&A \textbf{24}, 337 (1973)
\bibitem{spruit1995}
Spruit H. C., Stehle R., Papaloizou J. C. B., MNRAS \textbf{275},
1223 (1995)


\end{thebibliography}

%% % Non-BibTeX users please use
%% %

\end{document}